\documentclass[twocolumn,showpacs,preprintnumbers,amsmath,amssymb]{revtex4}

\usepackage{graphicx}
\usepackage{dcolumn}
\usepackage{bm}
\usepackage{latexsym}
\usepackage{graphics}
\usepackage{dcolumn}
\usepackage{epsfig}
\usepackage{amssymb} 
\usepackage{amsmath} 
\usepackage{rotate} 
\usepackage{color} 

\newcommand{\gammap}{\dot{\gamma}}

\newcommand{\dd}{\hbox{\rm d}}

\newcommand{\Rr}{R_{\scriptscriptstyle 1}}

\newcommand{\cz}{c_{\scriptscriptstyle 0}}
\newcommand{\vy}{v_{\scriptscriptstyle y}}

\newcommand{\vz}{v_{\scriptscriptstyle z}}
\newcommand{\vrad}{v_{\scriptscriptstyle r}}
\newcommand{\vtheta}{v_{\scriptscriptstyle \theta}}
\begin{document}

\title{Shear-induced fractures and three-dimensional motions in an organogel}

\author{Pauline Grondin}
\affiliation{Universit\'e Bordeaux I, Centre de Recherche Paul Pascal, UPR CNRS 8641, Avenue Schweitzer, F-33600 Pessac, FRANCE} 
\author{S\'ebastien Manneville\footnote{Present address: Laboratoire de Physique - CNRS UMR5672,\\ENS Lyon, 46 all\'ee d'Italie, 69364 Lyon cedex 07, FRANCE}}
\affiliation{Universit\'e Bordeaux I, Centre de Recherche Paul Pascal, UPR CNRS 8641, Avenue Schweitzer, F-33600 Pessac, FRANCE} 
\author{Jean-Luc Pozzo}
\affiliation{ISM, Universit\'e Bordeaux I, UMR CNRS 5255, 351 cours de la lib\'eration, F-33405 Talence c\'edex, FRANCE}
\author{Annie Colin\footnote{Corresponding author: {\tt annie.colin-exterieur@eu.rhodia.com}}}
\affiliation{LOF, Universit\'e Bordeaux I, UMR CNRS-Rhodia-Bordeaux 1 5258, 178 Avenue Schweitzer, 
F-33608 Pessac c\'edex, FRANCE}

\date{\today}

\begin{abstract}
The flow behavior of a viscoelastic organogel is investigated using ultrasonic velocimetry combined with rheometry. Our gel presents a decreasing flow curve, {\it i.e.}, the measured stress decreases as a function of the applied shear rate. Strikingly, we note that the local flow curve calculated from the velocity profiles also exhibits a decreasing part. We attribute this regime to the presence of a fracturing process and three-dimensional motions in the bulk of the sample. 
\end{abstract}


\maketitle
\section{Introduction}
Soft glassy materials are disordered materials with very long relaxation times. They encompass many different systems such as foams, emulsions, granular systems, pastes, and gels.
The response of such systems to an external shear stress is characterized by two regimes:
for stresses below the yield stress $\sigma_0$ they lose their ability to flow, remain jammed, and respond elastically, whereas for stresses above $\sigma_0$ they flow as liquids \cite{Barnes:99}. 
Despite this macroscopic shared  behavior, local velocimetry experiments have revealed that 
the flow of jammed materials is not universal \cite{Becu:06}. 
Nonadhesive emulsions \cite{Salmon:03}, microgels, and pastes were 
shown to flow homogeneously in the vicinity 
of the yielding transition \cite{Meeker:04}.
In contrast, Laponite gels \cite{Pignon:96} or adhesive emulsions \cite{Becu:06} display shear localization as they go
through the yielding transition \cite{Coussot:02}. 

In some particular cases, the applied shear stress decreases as a function of the applied shear rate and the flow curve presents a minimum \cite{Chen:94, Coussot:93}.
In a pioneering work, Pignon {\it et al.} \cite{Pignon:96} have studied thixotropic colloidal suspensions
displaying such a behavior. They have measured the local 
strain by introducing a fine filament of colored suspension in the sample and following its
deformation.  
Four flow regimes corresponding to four different states of the strain field were defined by combining visualization techniques and rheometric measurements in cone-and-plate geometry. 
At small deformations, the gels have an elastic behavior corresponding to a homogeneous strain field. 
Above a critical strain, the gel is deformed as a function of the applied shear rate. 
At very low shear rates, the thickness of the 
sheared layer is of the order of the size of the clay particles and the shear stress decreases 
as a function of the apparent shear rate. In the intermediate shear rate range, 
the size of the fluid layer increases and the shear stress remains constant as
the apparent shear rate increases. At high shear rates, the flow becomes homogeneous. 
This points out that the global flow curve does not reflect the local rheological behavior and raises the question of the origin of decreasing flow curves.

In this work, we revisit this problem using an organogel and ultrasonic velocimetry in Couette geometry \cite{Manneville:04} combined with standard rheometry. We show that three-dimensional motions and fractures occur in the decreasing part of the flow curve. This limits the possibility to etablish a valid flow curve even by using local tools such as local velocimetry. Indeed,  it is not possible to induce pure shear flow (even locally) in this system. 
This paper is organized as follows. Section II introduces the organogel under study, a mixture of an organic solvent and a low molecular weight gelator. We describe the synthesis of this system and study its structure as well as its basic linear rheological properties. In Section III, the flow properties of this thixotropic gel are addressed. We first use nonlinear rheology alone to access the flow curve of our system. Then the ultrasonic setup for measuring velocity profiles under shear is briefly recalled. Finally, evidences for fractures, three-dimensional motions, and a multivalued, decreasing local flow curve are provided and discussed.

\section{The organogel under study: synthesis, structure, and linear rheology}

The present work is devoted to the study of an ``organogel.'' Organogels are thermoreversible,
viscoeleastic materials consisting of low molecular weight gelators
(LMWG) self-assembled in an organic solvent 
into complex three-dimensional
structures. At high temperature, LMWG are quite soluble
in organic solvents and their solutions are liquid \cite{Terech:97,Weiss:00}. 
At low temperature, multiple nonconvalent interactions such as
hydrogen bonding, donor-acceptor, and hydrophobic
interactions between the organogel building blocks give
birth to a three-dimensional network of entangled and
connected fibers. This process is reversible, {\it i.e} the
gel can be melt again at high temperature and reformed
under cooling.

In recent years, these organogels have been
the subject of increasing interest mainly due to the
fundamental questions raised by the gel structures and
assembly mechanisms \cite{Kellog:93,Sobna:97,Weiss:00b,Lescanne:03,Lortie:02}. In this paper, we focus on their rheological properties. 
A mixture of N-5-hydroxypentyl-undecanamide (N5HU) in toluene has been studied. 
We first describe briefly the synthesis of N5HU and the gel formulation. Then the gel structure is characterized together with its linear rheological properties. 

\subsection{Synthesis of N-5-hydroxypentyl-undecanamide}

\begin{figure}
\begin{center}
\scalebox{0.4}{\includegraphics{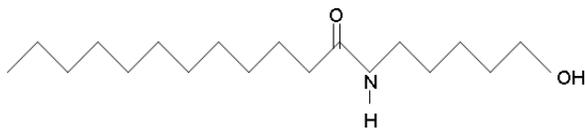}}
\end{center}
\caption{The N-5-hydroxypentyl-undecanamide molecule.}
\label{molecule1}
\end{figure}

The gelator molecule used in this work, namely N-5-hydroxypentyl-undecanamide (N5HU) 
is shown in figure~\ref{molecule1}.
The synthesis of N5HU  was previously described in Ref.~\cite{Kato:88}. Lauryl chloride (1 mol equiv) is added dropwise at 0$^\circ$C to a solution of 5-aminopentanol (Sigma) (1 mol equiv) in 40~mL of 0.2~N NaOH (Aldrich). 
The reaction is stirred for one night at room temperature. The precipitate is filtered 
and dissolved in dichloromethane
(Aldrich). The organic layer is washed several times
with water and dried over Na$_2$SO$_4$ (Sigma). After evaporation,
the white solid is triturated with pentane, filtered, and dried
under a vacuum. Toluene (Aldrich) is used as received. 
Purity of the sample was checked by
1H~NMR and IR spectroscopy.

This molecule is derived from previous studies on N-acyl-1, $\omega$-amino acid based gelators \cite{Terech:97b} and on N-3-hydroxypentyl-undecanamide~\cite{Lescanne:04}. It forms a gel in various fluids such as toluene and some mixtures of silicone oil in toluene
and of pentanol in dodecane. In the following, we present a study of gels of N5HU in toluene. 

\subsection{Gel formulation and phase diagram} 

To obtain a gel, heterogenous mixtures of various concentrations of gelator N5HU
(from 1 to 4.7 wt~$\%$) in toluene were heated (at about 70$^\circ$C) until clear liquids were obtained. The homogenous warmed solutions were then cooled rapidly at a given 
temperature thanks to a temperature-controlled bath.

\begin{figure}
\begin{center}
\includegraphics{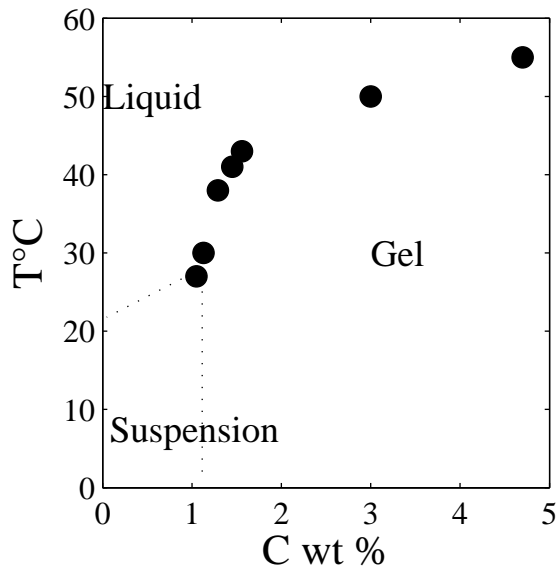}
\end{center}
\caption{Temperature-composition phase diagram for the N5HU-toluene system.}
\label{diag}
\end{figure}

Figure~\ref{diag} shows the phase diagram obtained by studying 
the transition temperature between the liquid and gel phase for 
the different samples. The procedure used to determine the transition temperature is rather crude.
The solution is prepared in a tube which is held upside down once the final temperature is reached. If the sample flows, it is called a liquid, whereas if it sustains its own weight, it is called a gel. For concentrations above 1.2~wt~$\%$, gels are obtained at low temperature and become
liquid at high temperature. The
gel can be melt again at high temperature and reformed
under cooling, so that the gelification process is reversible. The melting temperature and the gelification temperature differ by less than 1$^\circ$C.  For concentrations lower than 1.2~wt~$\%$
of N5HU, a suspension is observed at low temperature.

In order to characterize the structure of the gel and to understand why a gel is observed at low temperature, microscopic observations and linear rheology experiments have been 
performed and are presented in the next sections.

\subsection{Microscopic observations}

\begin{figure}
\begin{center}
\scalebox{1}{\includegraphics {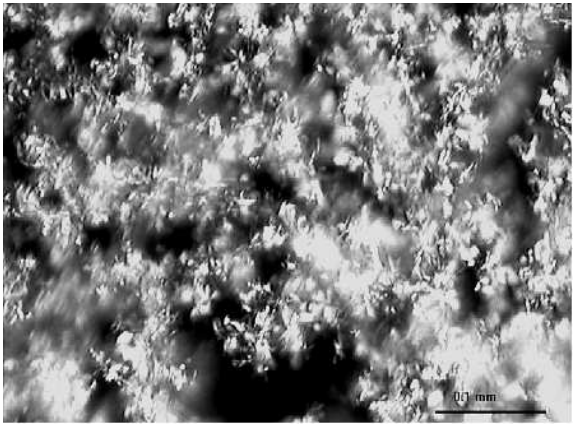}}
\end{center}
\caption{Optical microscopy picture of a 1.2~wt~$\%$ N5HU-toluene gel formed at a quench temperature of 5$^\circ$C. The picture was taken just after the quench. The scale bar represents 100~$\mu$m. }
\label{mgel1}
\end{figure}

\begin{figure}
\begin{center}
\includegraphics {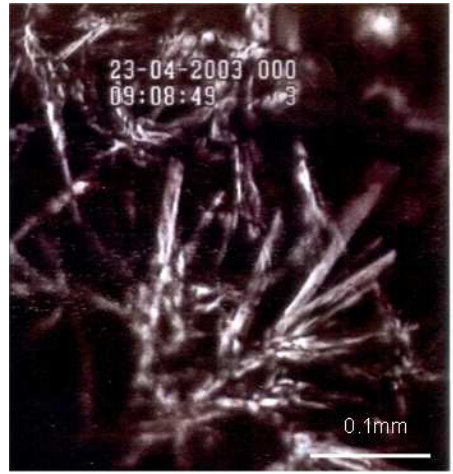}
\end{center}
\caption{Optical microscopy picture of a 1.2~wt
$\%$ N5HU-toluene gel at 5$^\circ$C one week after the
quench. The scale bar represents 100~$\mu$m.}
\label{vieuxgel}
\end{figure}

A solution of N5HU at 1.2~wt $\%$ in toluene is heated at 65$^\circ$C and 
cooled down to 5$^\circ$C at a cooling rate of 20$^\circ$C$/$min. 
Microscopic pictures of the resulting gel taken immediately after the quench and one week later are displayed on Figs.~\ref{mgel1} and \ref{vieuxgel} respectively.

This qualitative study leads to the following conclusions. 
The gel is  formed of large aggregates composed of interconnected branches or
fibers that have grown from the same center. The size of the fibers ranges from 20 to 200~$\mu$m. These aggregates form a network that is responsible for the elastic behavior of the
gel. Figure~\ref{vieuxgel} shows that the initial gel state is metastable and that the
final state (after 1 week) is a suspension of independent
and unconnected rods which has lost its elastic properties. With time some parts of the primary gel fibers melt and reform at other places with a different morphology. The mean diameter of the fibers increases, the new stucture is much less branched, and Oswald rippening seems to occur and to smooth out the stucture. This ageing behavior is similar to the one observed by Lescanne and coworkers on N-3-hydroxypentyl-undecanamide-toluene gels~\cite{Lescanne:04}.

Such an evolution is very slow, which 
allows us to consider that the structure does not vary significantly in the first five hours. All our experiments will deal with fresh samples and will be started a few minutes after the quench. Note that in the following, the rheological experiments will be performed on gels with 
higher N5HU concentrations. We shall assume that the structure involved in these systems 
is similar to the one presented above for lower concentrations. Indeed, 
due to the turbidity of the sample, we have been 
unable to obtain optical pictures for more concentrated systems.

\subsection{Linear rheology}

\begin{figure}
\begin{center}
\includegraphics{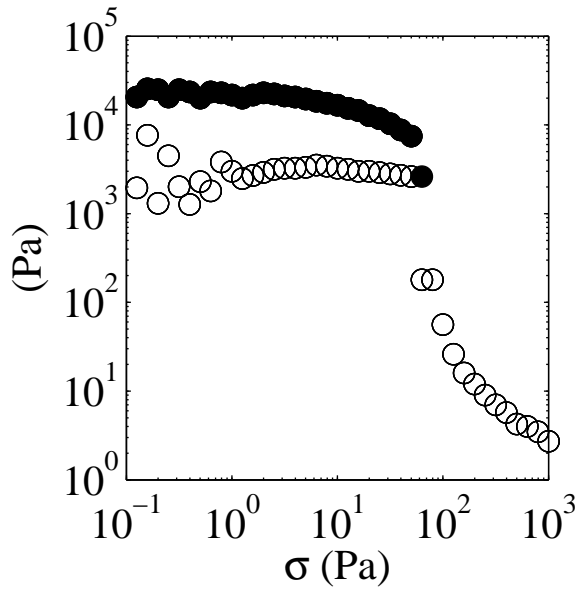}
\end{center}
\caption{Viscoelastic moduli $G'$ ($\bullet$) and $G''$ ($\circ$) as a function of the amplitude $\sigma$ of the applied shear stress for a 4.7~wt~\% N5HU-toluene gel. The frequency of the stress oscillations is 1~Hz.}
\label{dl5hu}
\end{figure}

To better characterize the nature of the interconnection between the aggregates, rheological 
experiments have been performed using a stress-controlled rheometer (TA Instruments AR 1000N) 
equipped with a cone-and-plate geometry made of aluminum (40 mm diameter, 4$^\circ$ cone angle, and 56 $\mu$m gap truncation). 
To avoid solvent evaporation, a solvent trap was placed above the cone. 
A striated cone-and-plate geometry was used to minimize wall slip. 
A heated solution (at around 65$^\circ$C) of 4.7~wt~$\%$ of N5HU in toluene is loaded 
in the shearing gap and cooled down to 
5$^\circ$C using the Peltier plate of the rheometer. A gel is obtained and the rheological experiment is started
10 minutes after loading the sample. In order to define the so-called linear regime,
the sample is first submitted to a periodic stress oscillating at a frequency of 1~Hz and with an amplitude $\sigma$ that increases exponentially with time. 
During this experiment, the storage modulus $G'$ and the loss modulus $G''$ 
are recorded as a function of the stress amplitude and plotted in Fig.~\ref{dl5hu}. 

At low stress values, {\it i.e.}, in the linear regime, both dynamic moduli are independent of the stress amplitude $\sigma$ and reveal the properties 
of the unpertubed gel: $G'$ is about 10 times larger than $G''$ which reflects the dominant character of the material. Up to large experimental scatter on $G''$ below 1~Pa, both moduli remain roughly constant
 below a critical shear stress of about 3~Pa, which can be taken as the limit of the linear domain. 
 Between $\sigma=3$~Pa and 60~Pa, $G'$ slightly decreases as a function of the applied shear stress. Above $\sigma=60$~Pa, $G'$ drops dramatically while $G''$ decreases sharply.
This behavior is due to the partial breakup of the gel. Thus our N5HU gel behaves as a weak gel \cite{Ross:06}.

\begin{figure}
\begin{center}
\includegraphics{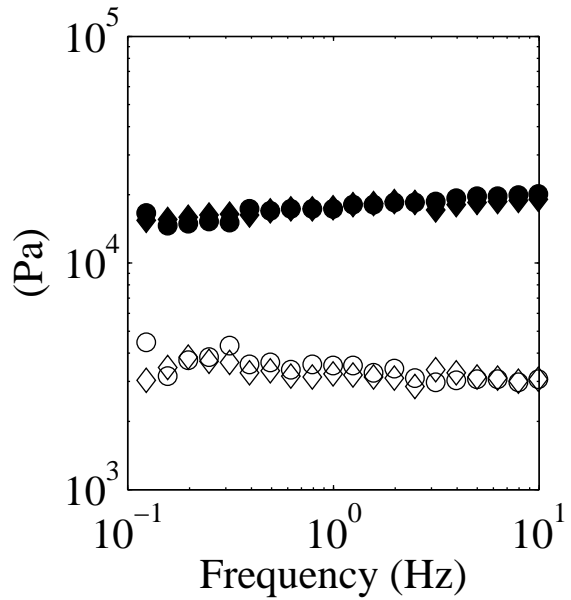}
\end{center}
\caption{Frequency sweep experiments for a 4.7~wt~\% N5HU-toluene gel: 
$G'$ ($\bullet$ and $\blacktriangledown$) and $G''$ ($\circ$ and $\diamond$) as a function of frequency. The amplitude of the applied shear stress is 2~Pa, which falls into the linear regime.}
\label{dln5hu}
\end{figure}

In a second set of experiments, $G'$ and $G''$ are measured as a function of the
 angular frequency (which is varied from 0.1 to 10~Hz) for a fresh sample of the N5HU toluene gel and a stress amplitude of 2 Pa. This stress falls into the linear regime so that the frequency sweep is nondestructive and the gel structure remains intact. Two different measurements performed on the same sample within 20 minutes are reported in Fig.~\ref{dln5hu} showing a good reproducibility of the experiment. A slight increase of the elastic modulus $G'$ as a
function of frequency is noticed (from $1.5\,10^4$~Pa at
0.1~Hz to $2\,10^4$~Pa at 10~Hz), which is consistent with
a viscoelastic behavior. On the other hand, the loss modulus is observed
to increase as the frequency decreases. Since $G''$ should reach zero at
zero frequency, a viscoelastic process occurs with a very
long characteristic time, corresponding to a very low
frequency value ($<0.1$~Hz). This long relaxation time may be attributed to
local rearrangements of the aggregates.

To conclude this study, it is important to note that reproducibility
is excellent on the same sample in the same day (as shown by Fig.~\ref{dln5hu})
but not as perfect when we compare the properties of two different
gel samples loaded and formed in the rheometer. The
values of $G'$ and $G''$ may vary by $40\%$ from one
experiment to another. This behavior is common to many
organogels. We believe that these discrepancies are related
to a dispersion in the aggregate number density from one
gel to another. Indeed, this parameter is very sensitive to cooling
velocities, impurities, and filling conditions of the geometry which
are difficult to control very precisely. After studying the main linear rheological properties of our organogel, we now focus on its nonlinear rheology. 

\section{Nonlinear flow behavior: evidence for fractures in the bulk and three-dimensional motions}

In this section, we study the nonlinear rheological response of our NH5U-toluene gels. 
We first address the question of the sensitivity of these gels to shear. Indeed, most organogels are very sensitive to shear stress which tends to expel the solvent from the three-dimensional network and to break the gel. When the mechanical excitation is stopped, the sample behaves as a suspension and has lost its elastic properties. Usually, the only way to recover a gel is to heat it and cool it down once again. Here we first present data showing that this is not the case for our NH5U-toluene gels and that recovery occurs spontaneously once shear is stopped. After discussing {\it global} nonlinear rheological properties, we present local velocity measurements using ultrasonic velocimetry and extract the {\it local} rheological flow curve. 

\subsection{Sensitivity of the gel to shear stress}

\begin{figure}
\begin{center}
\includegraphics{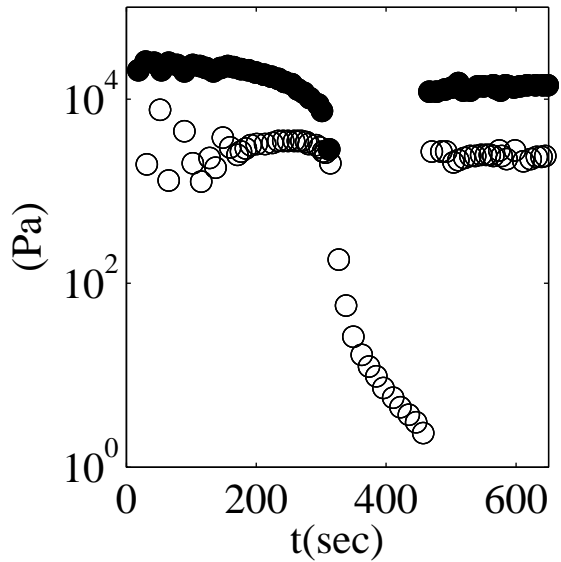}
\end{center}
\caption{Viscoelastic moduli $G'$ ($\bullet$) and $G''$ ($\circ$) as a function of time for a 4.7~wt~\% N5HU-toluene gel. The frequency of the stress oscillations is 1~Hz.
For $t<480$~s, the amplitude of the applied shear stress is increased from 0.1 Pa to 1000~Pa. For $t\ge 480$~s, a periodic stress of amplitude 1~Pa and frequency 1~Hz is applied. The gel recovers instantaneously its mechanical properties.}
\label{nonlimagn}
\end{figure}

To study the flow properties of the gel and its recovery, we go back to the experiment presented in Fig.~\ref{dl5hu} and performed on a 4.7~wt~$\%$ NH5U-toluene gel in the striated cone-and-plate geometry.
Figure~\ref{nonlimagn} presents the storage modulus $G'$ and the loss modulus $G''$ 
measured as a function of time. During the first 480~s, the stress amplitude $\sigma$ is increased exponentially in time from 0.1~Pa to 1000~Pa. These data are thus the same as those presented in Fig.~\ref{dl5hu} where they were shown as a function of the stress amplitude. The gel breaks at $t\simeq 300$~s for $\sigma\simeq 60$~Pa and then behaves like a viscous fluid ($G'=0$ for $300\lesssim t<480$~s). 

For $t\ge 480$~s, a small periodic stress of amplitude 1~Pa and frequency 1~Hz is applied. 
Instantaneously, the sample recovers its elastic behavior. This gel clearly exhibits 
a thixotropic behavior as defined in 1975 by the British Standards Institution \cite{Pignon:96}: the apparent viscosity decrease under stress is followed by a gradual recovery when stress is removed \cite{Cheng:87}.

\subsection{Global flow curve}

We now study the nonlinear behavior of our gel under constant applied shear rate. 
As previously, a 4.7~wt~$\%$ solution of NH5U in toluene is prepared and heated at 50$^\circ$C in order to totally dissolve the solid powdered sample. 
The hot solution is introduced between the two coaxial cylinders of a Couette 
geometry made of PVC (gap $e=0.5$~mm, inner radius $\Rr=24.5$~mm and height $H=30$~mm) and then cooled down to 5$^\circ$C with a cooling rate of about 20$^\circ$C/min using an icy water circulation around the outer cylinder. The reason for using a Couette geometry in this section is that we wish to further use ultrasonic velocimetry (which is currently available only in Couette geometry) in order to construct a local flow curve by combining global rheological data and local velocity measurements. The same behaviour was observed in cone-and-plate geometry even though the global flow curve may vary a little bit. 

\begin{figure}
\begin{center}
\includegraphics{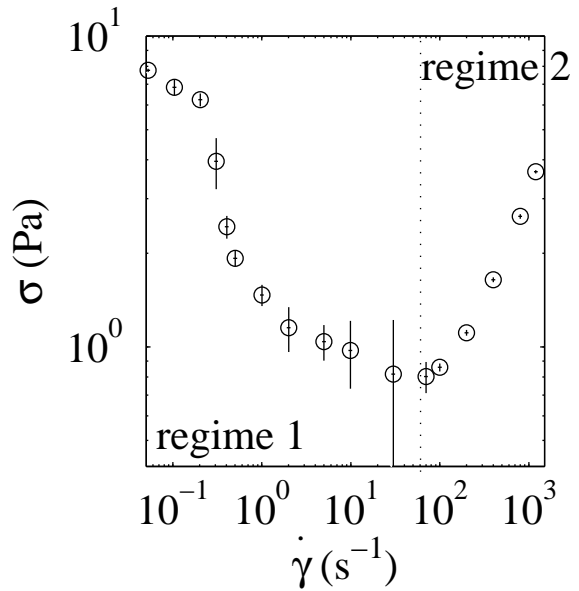}
\end{center}
\caption{Global flow curve: shear stress $\sigma$ as a function of the applied shear rate $\gammap$ for a 4.7~wt~\% N5HU-toluene gel. Two different regimes are evidenced below and above $\gammap\simeq 70$~s$^{-1}$.}
\label{flowcurve}
\end{figure}

Using the shear rate imposed mode of the AR1000N rheometer, a constant shear rate $\gammap$ is applied to the sample for 10~min and the shear stress $\sigma(t)$ is recorded as a function of time. This procedure is applied for $\gammap$ ranging from 0.05 to 1400 s$^{-1}$. Figure~\ref{flowcurve} presents the {\it global} flow curve $\sigma$ versus $\gammap$ (also referred to as the ``engineering'' flow curve in the literature) where the measured shear stress values are obtained by averaging $\sigma(t)$ over the last 300~s of each shear rate step. Waiting for 5~min before starting to average the rheological data ensures that steady state is reached. Errors bars correspond to the standard deviation of $\sigma(t)$ and indicate noisy or oscillatory signals.

The flow curve of Fig.~\ref{flowcurve} may easily be divided into two different regimes. First, for $0.05\le\gammap\lesssim 70$~s$^{-1}$ (regime 1), the
measured shear stress decreases as a function of the applied shear rate. In the case of pure shear flow, a negative slope in the flow curve cannot be observed experimentally since it corresponds to material instability. Thus the global flow curve of Fig.~\ref{flowcurve} strongly suggests the presence of inhomogeneous flows in regime 1.
Second, for $\gammap\gtrsim 70$~s$^{-1}$ (regime 2), the measured shear stress increases as a function of the applied shear rate, consistently with a low-viscosity liquid behavior. 
Note that, if we repeat the experiments after shearing the system at a shear rate of 800~s$^{-1}$ for 2~min and stopping the shear for 5~min, we get the same rheological behaviour. This confirms the thixotropic nature of our gel.
To better analyze this striking flow curve, we have performed local velocity experiments simultaneously to nonlinear rheological measurements. 

\subsection{Experimental setup for ultrasonic velocimetry}
\label{expsetup}

\begin{figure}
\begin{center}
\scalebox{0.8}{\includegraphics{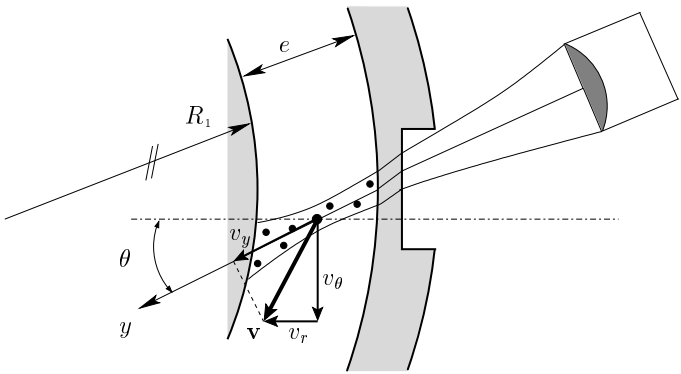}}
\end{center}
\caption{Experimental setup for ultrasonic velocimetry under shear in Couette geometry. The notations are defined in the text.}
\label{f.setup}
\end{figure}

Our rheo-ultrasonic setup allows the recording of velocity
profiles across the gap of a Couette cell simultaneously to rheological data with a spatial resolution of about 40 $\mu$m. The time needed to record one velocity profile ranges between $0.02$ and $20$~s, depending upon the value of the applied shear rate.
This setup has been described at length in Ref.~\cite{Manneville:04} and only its
main characteristics are recalled here. It is built around the stress-controlled rheometer (TA Instruments AR1000N) used so far and equipped with
the Couette cell described above. The whole cell is surrounded by water whose temperature is kept constant to within $\pm$0.1$^\circ$C. 
A piezo-polymer transducer of central frequency 36~MHz generating focused ultrasonic pulses
is immersed in water in front of the stator, as sketched in Fig.~\ref{f.setup}.		
The transducer is controlled by a pulser-receiver unit
(Panametrics 5900PR) which generates 220~V pulses
with a rise time of about 1~ns. The pulse repetition frequency $f_{PRF}$ is tunable from 0 to 20~kHz. Backscattered (BS) signals are sampled at 500~MHz, stored on a PCI digitizer with 8~Mb on-board memory (Acqiris DP235), and later transferred
to the host computer for post-processing.

\begin{figure}
\begin{center}
\includegraphics{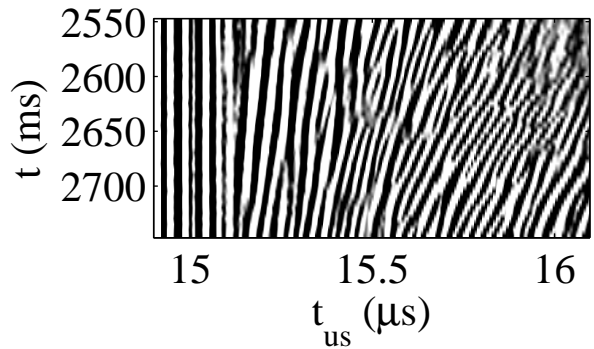}
\end{center}
\caption{BS signals corresponding to 40 pulses sent every $\tau=6.7$~ms in a Newtonian latex suspension sheared at $\gammap=10$~s$^{-1}$. The normalized pressure $p/p_0$ is coded in gray
levels. The horizontal axis is the (fast) ultrasonic time
$t_{us}$ and the vertical axis is the (slow) pulse time $t$. The gap lies between $t_{us}\simeq 15.1~\mu$s and $t_{us}\simeq 16.0~\mu$s.}
\label{billes}
\end{figure}

Under the assumption of single scattering, the signal received at time 
$t_{us}$ can be interpreted as interferences coming from scatterers located at position  
$y=\cz t_{us}/2$, where $\cz$ is the sound speed in the fluid and $y$ is the distance from
the transducer along the ultrasonic beam (see Fig.~\ref{f.setup}). 
When the sample is submitted to a shear flow, BS signals move along with the scatterers. 
Two successive pulses separated by a time interval $\tau=1/f_{PRF}$ lead 
to two similar backscattered signals
that are shifted in time. Figure~\ref{billes} 
shows forty successive BS signals recorded
in a sheared Newtonian latex suspension with a pulse frequency repetition $f_{PRF}=150$~Hz. The amplitude of the BS signals is coded in gray levels and the vertical axis $t$ corresponds to the time at which each pulse is sent.
The ultrasonic pulses leave the stator and enter the gap
at $t_{us}\simeq 15.1~\mu$s and the rotor
 position corresponds to $t_{us}\simeq 16.0~\mu$s.
The slopes of the traces left by the echoes in this two-dimensional 
($t_{us}$, $t$) diagram are inversely proportional to the local velocities. 
Thus, the signature of shear is rather clear: velocities increase from
the stator, where the scatterers remain almost fixed, to the
rotor. Moreover, the presence of moving echoes for
$t_{us}>16~\mu$s is not surprising: these echoes simply correspond to
scattering of the wave reflected on the rotor. One can see
that the whole range $t_{us}=15.1$--16.0~$\mu$s gets covered
by some significant speckle signal in about 20~ms in that
case.

From these BS signals, one can relate the time shift $\delta t(t_{us})$ between 
two echoes corresponding to two successive pulses and received at time $t_{us}$ to the displacement of the 
scatterers $\delta y(y)=\cz~ \delta t(t_{us})/2$ at position $y$. 
The velocity profile $\vy(y)$ ({\it i.e.}, the $y$ component of the velocity vector 
$\mathbf{v}$) is then simply given by $\vy(y) =\delta y(y)/\tau$.
A cross-correlation algorithm is used to estimate the time shift $\delta t$ as a 
function of $t_{us}$ and the sound speed $\cz$ is measured independently (see Ref.~\cite{Manneville:04} for more details).
In order to get a non-zero projection of the velocity along the acoustic axis, ultrasonic
pulses enter the gap with an angle $\theta\simeq 15^\circ$ as shown in Fig.~\ref{f.setup}. 

Using the notations of Fig.~\ref{f.setup}, the projection $\vy$ of the velocity vector $\mathbf{v}$ along the $y$-axis
in Couette geometry is the sum of the projections of
the radial component $\vrad$ and of the tangential component $\vtheta$ of $\mathbf{v}$:
\begin{equation}
\vy=\vtheta \sin\theta +\vrad \cos\theta\,.
\label{e.vy}
\end{equation}
Assuming the flow field to be purely tangential,
one gets:
\begin{equation}
\vtheta(x)=\frac{\vy(y)}{\sin\theta}\equiv v(x)\,,
\label{e.defvx}
\end{equation}
where $x$ is the radial position given by $x=e-y\cos\theta$.
The determination of $v(x)$ thus requires a precise determination of the angle $\theta$, which is achieved
through a careful calibration procedure using a Newtonian fluid \cite{Manneville:04}.
If the radial component $\vrad$ is non-zero, the term $\vrad\cos\theta$ comes into play
in Eq.~(\ref{e.vy}), so that the definition of $v(x)$ given in Eq.~(\ref{e.defvx}) leads to 
\begin{equation}
v(x)=\vtheta(x)+\frac{\vrad(x)}{\tan\theta}\,.
\label{e.modvx}
\end{equation}
Therefore, the interpretation of the experimental velocity profiles $v(x)$
may become problematic whenever the flow is not purely tangential. 
It should also be mentioned that our setup is insensitive to any contribution of the vertical component $\vz$ to the
velocity vector ({\it i.e.}, in the vorticity direction $z$), since its projection
along the acoustic axis is always zero.

In principle, a velocity profile could be obtained by cross-correlating only two succesive BS signals. However, in practice the speckle amplitude is never uniform. 
Locally, destructive interferences or the absence of scatterers may lead to signal levels too small to be well analyzed. To recover a full velocity profile, some averaging is then performed over several successive cross-correlations. 
Typically, on a stationary flow, 100 cross correlations are required to get a standard deviation of the velocity measurements of less than $6\%$ \cite{Manneville:04}. 
In the following, we will present both time-averaged data and ``instantaneous'' measurements {\it i.e.} data obtained using only two successive cross-correlation.

Finally, thanks to their specific microstructure made of rigid aggregates whose size is comparable to the acoustic wavelength used for velocimetry, our NH5U gels naturally scatter ultrasonic waves in the single scattering regime. Thus it was not necessary to seed the gel with scatterers to perform velocity measurements. The next section displays and analyzes the time-averaged data recorded in the NH5U organogel.

\subsection{Time-averaged velocity profiles}

In this section, we focus on the time-averaged velocity profiles recorded using ultrasonic velocimetry once rheological variables have reached steady state for each shear rate step along the flow curve of Fig.~\ref{flowcurve}. To this aim the local velocity measurements $v(x)$ were averaged over 6 to 10 series of 1000 pulses (except for shear rates smaller than 0.3~s$^{-1}$ where 2 to 5 series of 500 pulses were used in order to fit within the last 300~s of each shear rate step). 15 measurement points are obtained over the gap width $e=0.5$~mm correponding to a spatial resolution of 33~$\mu$m. The standard deviation of these measurements then reflects the occurence of temporal velocity fluctuations.

\subsubsection{Low shear regime (regime 1)}

\begin{figure}
\includegraphics{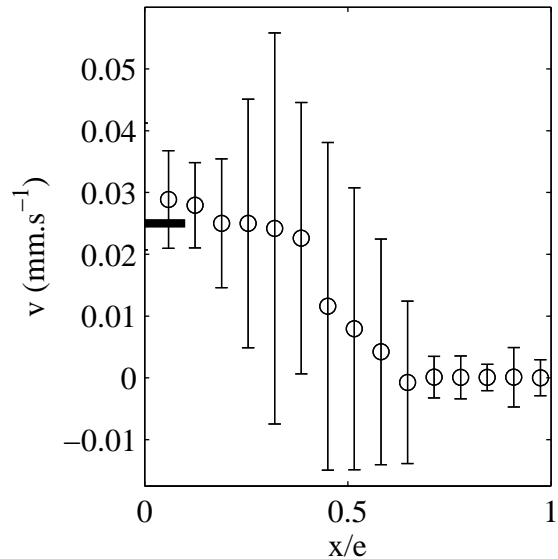}
\caption{Time-averaged velocity profile $v(x)$ as a function of the reduced coordinate $x/e$ for $\gammap=0.05$~s$^{-1}$. The horizontal bar on the left axis indicates the rotor velocity. No wall slip can be clearly evidenced and the measurements are very noisy. The profiles have been averaged on the last 100~s of the shear rate step at $\gammap=0.05$~s$^{-1}$ once rheological variables have reached steady state.}
\label{005v}
\end{figure}

\begin{figure}
\includegraphics{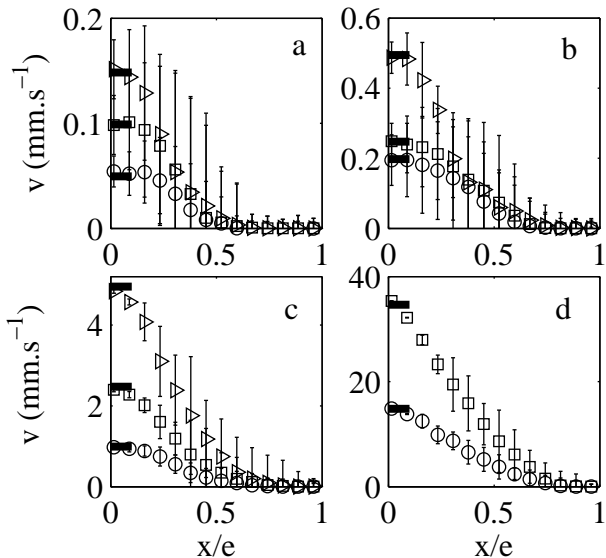}
\caption{Time-averaged velocity profiles $v(x)$ in regime 1 as a function of the reduced coordinate $x/e$. (a) $\gammap=0.1$ ($\circ$), 0.2 ($\square$), and 0.3~s$^{-1}$ ($\triangleright$). (b) $\gammap=0.4$ ($\circ$), 0.5 ($\square$), and 1~s$^{-1}$ ($\triangleright$). (c) $\gammap=0.2$ ($\circ$), 5 ($\square$), and 10~s$^{-1}$ ($\triangleright$). (d) $\gammap=30$ ($\circ$) and 70~s$^{-1}$ ($\square$). The horizontal bars on the left axis indicate the respective rotor velocities. No wall slip is evidenced.}
\label{zone1t}
\end{figure}

Figures~\ref{005v} and  \ref{zone1t} show the resulting time-averaged velocity profiles for shear rates ranging within the decreasing part of the flow curve (regime 1). By definition of $x$, $x=0$ ($x=e$ resp.) corresponds to the rotor (stator resp.). Three main features show up. We first note that for all applied shear rates, the velocity of the sample vanishes near the stator and reaches the velocity of the moving wall near the rotor. We conclude that no wall slip occurs for this gel. 

Second, the standard deviations of the velocity measurements are huge: they may reach 100$\%$ of the measured value at low applied shear rate (see errors bars on  Figs.~\ref{005v} and \ref{zone1t}). In some cases ``instantaneous'' negative values of the velocity are even recorded (see Fig.~\ref{005v}). The standard deviation decreases as a function of the applied shear rate. Such large error bars clearly reflect the occurence of strong temporal velocity fluctuations (since error bars of less than $3\%$ would be expected for a stationary flow with the same averaging \cite{Manneville:04}). We will come back to this point in details in Sect.~\ref{instant} below.

Third, in the bulk, three different zones (or ``shear bands'') can be clearly distinguished. (i) Close to the stator, the gel does not flow and behaves like a solid (see $x/e\in [0.55,1]$ in Fig.~\ref{005v} and $x/e\in [0.75,1]$ in Fig.~\ref{zone1t}). (ii) In the middle of the gap 
(see $x/e\in [0.4,0.55]$ in Fig.~\ref{005v}, $x/e\in [0.3,0.75]$ in Fig.~\ref{zone1t}(a), $x/e\in [0.2,0.75]$ in Fig.~\ref{zone1t}(b), $x/e\in [0.1,0.75]$ in Fig.~\ref{zone1t}(c), 
and $x/e\in [0.05,0.75]$ in Fig.~\ref{zone1t}(d)), the gel is flowing and bears a non-zero local shear rate. The value of this local shear rate increases as a function of the engineering applied shear rate. In the following, this zone will be called the ``viscous zone.'' (iii) Finally, close to the rotor (see $x/e\in [0,0.4]$ in Fig.~\ref{005v}, $x/e\in [0,0.3]$ in Fig.~\ref{zone1t}(a), $x/e\in [0,0.2]$ in Fig.~\ref{zone1t}(b), $x/e\in [0,0.1]$ in Fig.~\ref{zone1t}(c), and $x/e\in [0,0.05]$ in Fig.~\ref{zone1t}(d)), 
the gel remains unsheared. The size of this solid-like region located near the rotor decreases as the engineering applied shear rate is increased. At the end of regime 1 (see $\gammap=70$~s$^{-1}$ in Fig.~\ref{zone1t}(d)), the solid-like state near the rotor disappears and a flowing liquid coexists with a solid-like region near the stator.

\begin{figure}
\begin{center}
\includegraphics{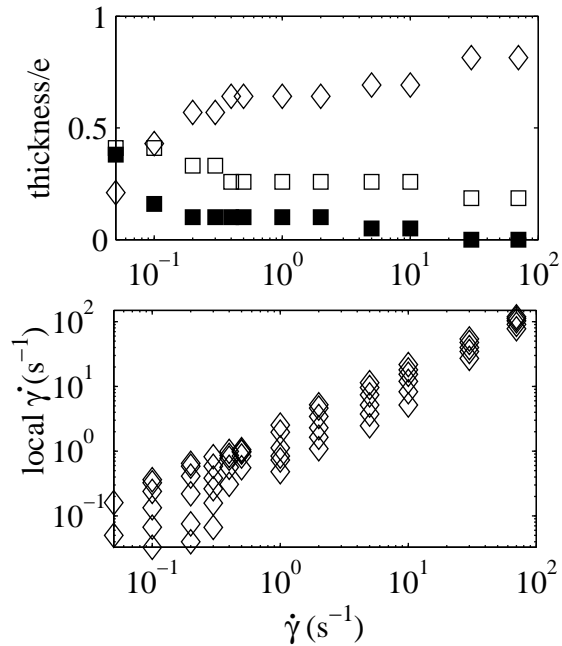}
\end{center}
\caption{Analysis of time-averaged velocity profiles in regime~1. Top: thickness of the shear bands divided by the gap width $e$ as a function of the engineering applied shear rate $\gammap$. $\square$ correspond to the solid-like state located near the stator, $\diamond$ to the middle zone, {\it i.e.}, to the viscous zone, and $\blacksquare$ to the solid-like state located near the rotor. Bottom: local shear rates borne by the sample in the viscous zone as a function of the engineering applied shear rate $\gammap$.}
\label{loc2}
\end{figure}

Figure~\ref{loc2} displays the thickness of the three shear bands and the local shear rates in the viscous zone as a function of the engineering applied shear rate in regime 1.
The local shear rate is extracted from the time-averaged velocity profiles using: 
\begin{equation}
\gammap(x)=-r\, \frac{\dd}{\dd r}\left(\frac{v(x)}{r}\right)\,,
\label{shear}
\end{equation}
where $r=\Rr+x$.
These data clearly reveal that the growth of the viscous zone occurs in the decreasing part of the engineering flow curve ({\it i.e.}, for $\gammap\le 70$~s$^{-1}$). 
This is not the case in clay gels where the growth of this region was shown to take place at a given shear stress \cite{Pignon:96}.
We also note that the flow behavior of this viscous zone is rather complex. Indeed for a given engineering applied shear rate, the local shear rate may vary by a factor of ten within the viscous zone. More precisely the local shear rate increases as a function of the reduced coordinate $x/e$: highest shear rates are evidenced near the solid-like state close to the stator. This result will be discussed in more details in the next section. 

\subsubsection{High shear regime (regime 2)}

\begin{figure}
\begin{center}
\includegraphics{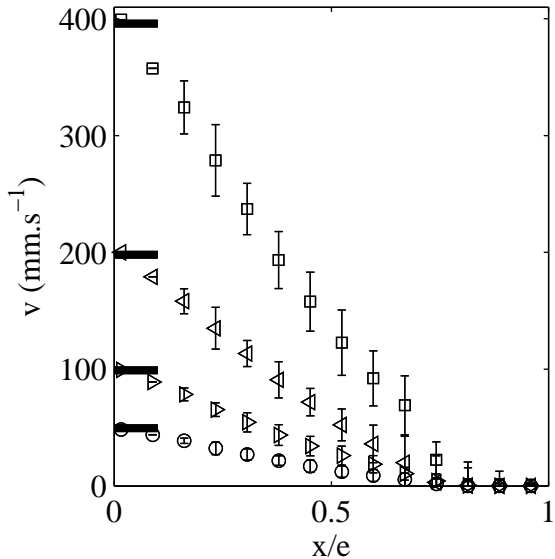}
\end{center}
\caption{Time-averaged velocity profiles $v(x)$ in regime 2 as a function of the reduced coordinate $x/e$. $\gammap=100$ ($\circ$), 200 ($\triangleright$), 400 ($\triangleleft$), and 800~s$^{-1}$ ($\square$). The horizontal bars on the left axis indicate the respective rotor velocities. No wall slip is evidenced. A solid-like region is detected near the stator even for the highest applied shear rates.} 
\label{zone2b}
\end{figure}

Time-averaged velocity profiles measured in regime 2 are shown in Fig.~\ref{zone2b}. 
In this regime, a flowing zone always coexists with a solid-like region near the stator. 
Increasing the applied shear rate increases the shear rate borne 
by the flowing zone, but almost does not change the size of the solid-like region near the stator. Error bars are still noticeable in regime 2 but their magnitudes are much less than those recorded at low shear rate.

\subsection{Local flow curve}
\label{locflowcurve}

To go deeper into the analysis of the flow behavior, 
one may extract a {\it local} flow curve from the combined velocity and rheological measurements \cite{Becu:06,Salmon:03,Huang:05}.
Indeed, the torque $\Gamma$ imposed on the moving cylinder by the rheometer yields 
the stress distribution $\sigma(x)$ across the Couette cell,
\begin{equation}
\sigma(x)=\frac{\Gamma}{2\pi H r^2}\,,
\end{equation}
and the velocity profile leads to the local shear rate $\gammap(x)$ according to eq.~(\ref{shear}).

\begin{figure}
\begin{center}
\includegraphics{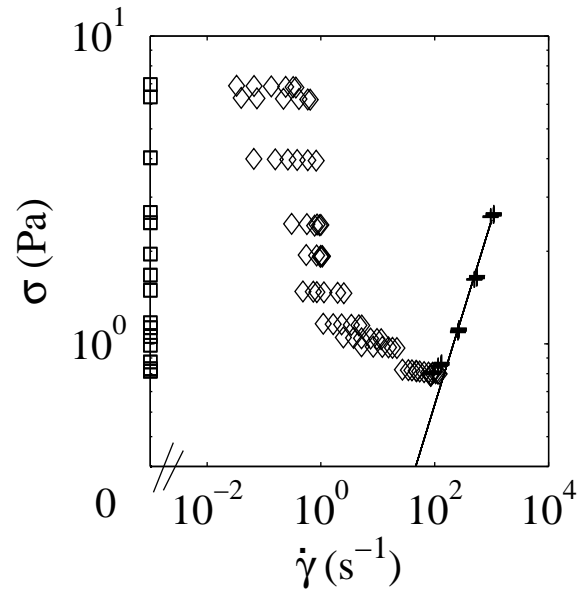}
\end{center}
\caption{Local flow curve $\sigma(x)$ versus $\gammap(x)$ inferred from rheological and local velocity measurements. Three different branches are evidenced: a solid-like state ($\square$), a viscous state ($\diamond$) in regime 1, and a low viscosity shear thinning fluid ($+$) in regime 2. The solid line corresponds to the shear thinning behavior $\sigma= A \gammap^n$ with $A=0.04 $ SI and $n=0.6$. }
\label{loc}
\end{figure}

The resulting $\sigma(x)$ versus $\gammap(x)$ data are plotted in Fig.~\ref{loc}.
The first striking observation is that the local flow curve is {\it multivalued}: for shear stresses ranging between 0.7 and 4~Pa, three different shear rates correspond to the same shear stress. The gel may thus adopt three different states for a given shear stress: (i) a solid-like state ($\square$), (ii) a viscous state ($\diamond$), and (iii) a low viscosity shear thinning state ($+$). State (ii) is observed in regime 1 and corresponds to the viscous zone in the velocity profiles of Fig.~\ref{zone1t}. The shear thinning local behavior (iii) is observed in regime 2 (see solid line in Fig.~\ref{loc}) and corresponds to the flowing region in Fig.~\ref{zone2b}.

Second, we note that flowing samples are observed for shear stresses below the upper limit of the linear regime $\sigma\simeq 3$~Pa. This simply means that during the experiments under oscillating shear stress, only the solid-like state was probed. Moreover, states (ii) and (iii) are obtained after a fracturing process which probably requires a certain amount of strain and thus may not have enough time to develop under stress oscillations. 

Finally the most  important result is the presence of a {\it decreasing} branch in the local flow curve ($\diamond$ and $\circ$ in Fig.~\ref{loc}). In regime 1, the shear stress decreases from 7 to 0.7~Pa  while the applied shear rate is increased from 0.05 to 70~s$^{-1}$. Usually the apparent shear viscosity is defined as the ratio between $\sigma(x)$ and $\gammap(x)$. Decreasing branches in  the flow curve are mechanicaly unstable. It is thus striking to get sound measurements in this zone of the flow curve. 
Let us recall that the velocity profiles display very large errors bars in this range of applied shear rates, reflecting important temporal fluctuations. In order to get further insight on this decreasing branch, it is thus necessary to focus on the ``instantaneous'' velocity profiles and to study in details the temporal behaviour of the local velocities. The next section is devoted to this point. 


\subsection{Analysis of backscattered ultrasonic signals and ``instantaneous'' velocity signals}
\label{instant}

\begin{figure}
\begin{center}
\includegraphics{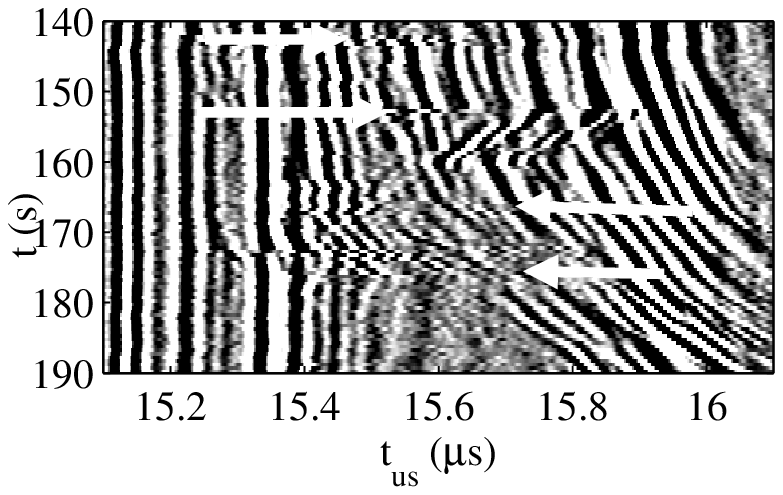}
\includegraphics{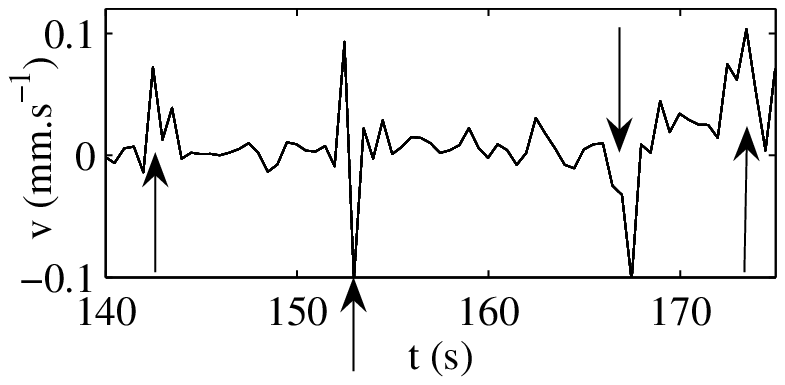}
\end{center}
\caption{Top: BS pressure signals recorded in a 4.7~wt~\% NH5U-toluene gel under an applied shear rate $\gammap=0.05$~s$^{-1}$ with $\tau=0.5$~s. Arrows highlight discontinuities in the BS signals versus the pulse time $t$, which we interpret as the signature of fracture-like events. Bottom: ``instantaneous'' velocity extracted from the above data as a function of time at $x=0.33$~mm (corresponding to $t_{us}=15.70~\mu$s). Arrows indicate the fracture events.}
\label{vitlocal005}
\end{figure}

In order to better understand the decreasing part of the local flow curve, we go back to the backscattered ultrasonic signals recorded in regime 1. 
Figure~\ref{vitlocal005}(top) displays backscattered ultrasonic signals recorded in the 4.7~wt~\% NH5U-toluene gel during the experiment of Fig.~\ref{flowcurve} at the very beginning of the shear rate step at $\gammap=0.05$~s$^{-1}$. More precisely, shear is applied at $t\simeq 140$~s and 100 successive signals with $f_{PRF}=2$~Hz are shown. Once shear is started, a flowing zone is nucleated in the middle 
of the gap after a few seconds and grows in time until steady state is achieved. The flow is thus clearly inhomogeneous.

\begin{figure}
\begin{center}
\includegraphics{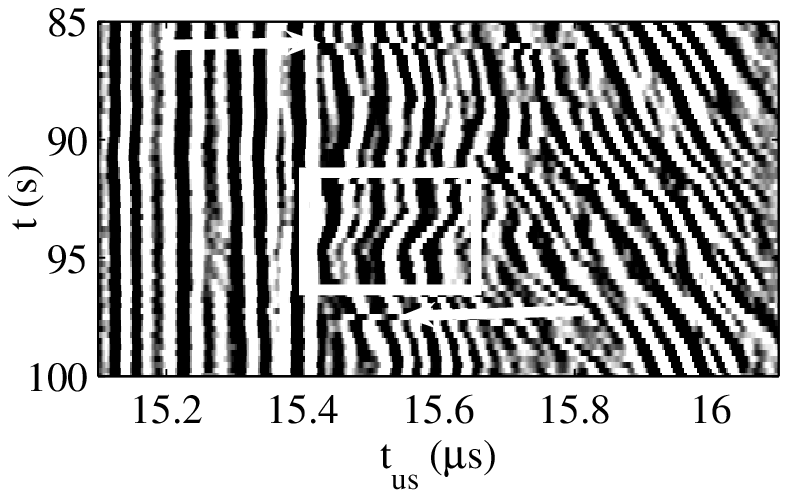}
\includegraphics{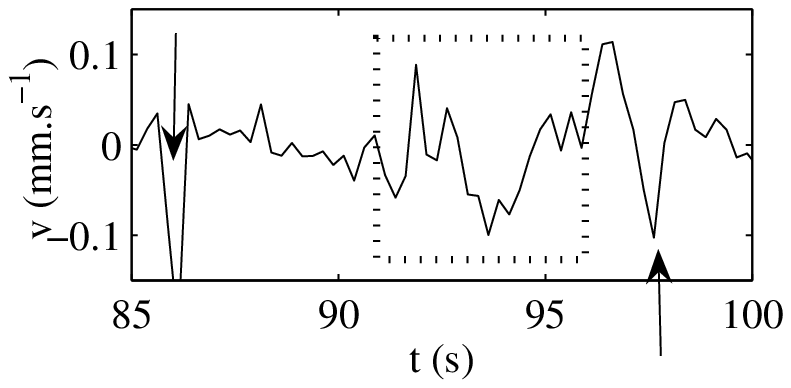}
\end{center}
\caption{Top: BS pressure signals recorded in a 4.7~wt~\% NH5U-toluene gel under an applied shear rate $\gammap=0.1$~s$^{-1}$ with $\tau=0.25$~ms. Arrows highlight discontinuities in the BS signals versus the pulse time $t$, which we interpret as the signature of fracture-like events. A typical zone of three-dimensional motion is surrounded by a white frame. Arrows indicate fracture-like events. Bottom: `instantaneous'' velocity extracted from the above data as a function of time at $x=0.2$~mm (corresponding to $t_{us}=15.46~\mu$s). Arrows indicate the fracture events and the dotted frame points to negative velocity data on ``slow'' time scales, which are indicative of three-dimensional motion.}
\label{fracture3D}
\end{figure}

A second very striking feature of the BS signals at $\gammap=0.05$~s$^{-1}$ is that the traces in Fig.~\ref{vitlocal005}(top) present sudden discontinuities from one pulse to another, {\it i.e.}, as a function of the pulse time $t$ (see arrows in Fig.~\ref{vitlocal005}). For the sake of clarity, we have also plotted the temporal evolution of the ``instantaneous'' velocity at a given point in the gap 
($x=0.29$~mm, which corresponds to $t_{us}=15.7~\mu$s) in Fig.~\ref{vitlocal005}(bottom). Such ``instantaneous'' measurements were obtained by cross-correlating two successive signals without further averaging cross-correlations. Although this procedure leads to rather noisy velocity signals, very fast variations of the ``instantaneous'' velocity (which in this case becomes negative) are revealed, corresponding to the sudden discontinuties pointed out by arrows in Fig.~\ref{vitlocal005}(top). Moreover the fact that these discontinuities do not span over the full range of ultrasonic times $t_{us}$ is an indication that ``fast'' events occur that are localized in both space and time. The time scale for such events is smaller than the time interval $\tau$ between two pulses. Even if more quantitative experiments would be needed  (in particular on much shorter time scales) in order to draw definite conclusions, we interpret these events as fractures in the bulk of the fluid. Such a behavior persists when steady state is achieved. Figure~\ref{fracture3D} shows that discontinuities in BS signals are also found for $\gammap=0.1$~s$^{-1}$.

\begin{figure}
\begin{center}
\includegraphics{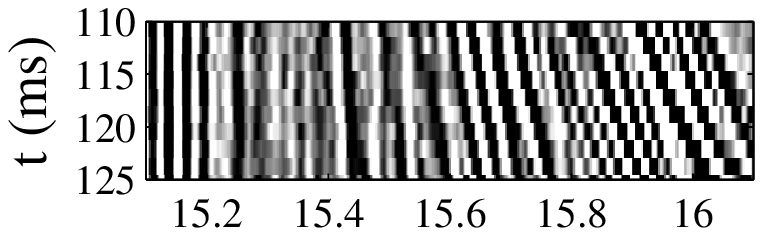}
\includegraphics{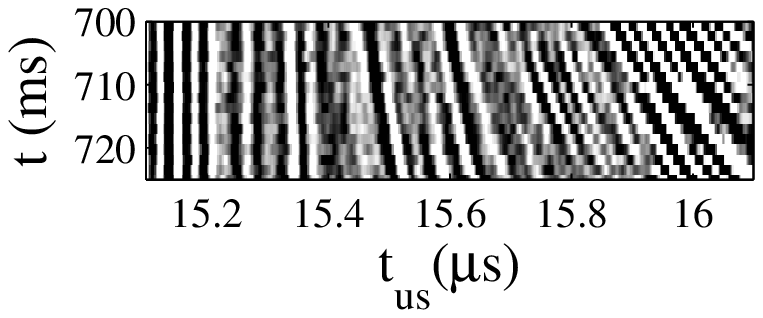}
\end{center}
\caption{BS pressure signals recorded in a 4.7~wt~\% NH5U-toluene gel under an applied shear rate $\gammap=200$~s$^{-1}$ with $\tau=0.25$~ms. Each subfigure displays a series of 20 successive signals. No three-dimensional motions or fracture-like events are evidenced.}
\label{fracture200}
\end{figure}

The third important point to notice is that the local slopes of the BS signals present temporal variations on ``slow'' time scales (typically a few time intervals $\tau$ between two pulses). In some cases, the slopes do not only vary but also change sign even when steady state is reached. Such events are highlighted for $\gammap=0.1$~s$^{-1}$ by white frames in Fig.~\ref{fracture3D}. They are also found at other shear rates in regime 1 (see $t\simeq 150$--160~s in Fig.~\ref{vitlocal005}(top) and Fig.~\ref{fracture200}). This apparent ``reverse'' motion of the ultrasonic echoes can be attributed either to some rotational motion of the N5HU aggregates or to some significant radial component of the velocity field. Although we cannot discriminate between these two mechanisms through the present experiments alone, our observations constitute a clear evidence for three-dimensional features of the displacement field with characteristic times of a few seconds. As explained above in Sect.~\ref{expsetup}, the consequence of the three-dimensional nature of the flow is the possibility to measure apparent negative velocities on rather long time scales. This is illustrated in Fig.~\ref{fracture3D}(bottom) where the velocity remains negative for $t\simeq 92$--94~s.

\begin{figure}
\begin{center}
\includegraphics{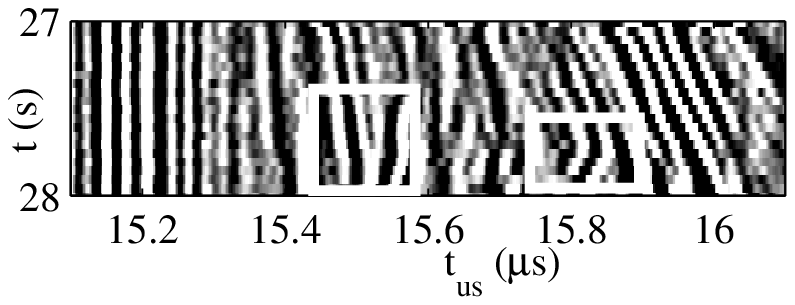}
\includegraphics{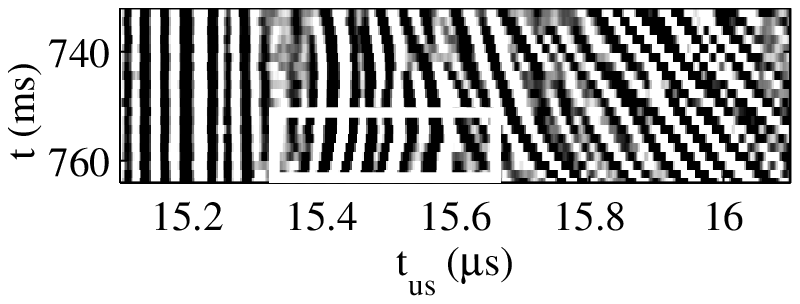}
\end{center}
\caption{BS pressure signals recorded in a 4.7~wt~\% NH5U-toluene gel under an applied shear rate $\gammap=1$~s$^{-1}$ with $\tau=50$~ms (top) and $\gammap=70$~s$^{-1}$ with $\tau=1.4$~ms (bottom). Each subfigure displays a series of 20 successive signals. Some typical zones of three-dimensional motion are surrounded by a white frame.}
\label{fracture100}
\end{figure}

Three-dimensional motions and fractures persist throughout regime 1 but these events become more rare and weaker as the shear rate is increased. Figure~\ref{fracture100} shows more examples of three-dimensional flow features recorded in regime 1 at $\gammap=1$ and 70~s$^{-1}$. However no fracture-like events can be detected on these BS signals. As illustrated by Fig.~\ref{fracture200}, three-dimensional behaviors as well as fractures completely disappear in regime 2.

This qualitative analysis allows us to better understand the huge errors bars reported in the time-averaged velocity profiles. As suspected, these error bars originate from very large temporal fluctuations in the flow field. We have shown that these fluctuations arise either from ``fast'' fracture-like events (on time scales smaller than $\tau$) or from ``slow'' three-dimensional motions (on time scales of a few $\tau$). Such features are not found at higher shear rates. We believe that such fluctuations cancel out when time-averaged data are considered so that the velocity profiles of Figs.~\ref{005v} and {zone1t} may still be interpreted in terms of tangential velocity.

From these observations we may propose an explanation for the negative slope of the local flow curve reported in Sect.~\ref{locflowcurve}. Indeed, fractures and three-dimensional flow features lead to enhanced friction losses in the sample. Since their number and amplitude were shown to decrease as the shear rate is increased in regime 1, the loss of energy becomes smaller, which could lead to a decrease of the shear stress. Note, however that as we are unable to submit the gel to pure shear flow, it is thus not possible to identify $\sigma(x)/\gammap(x)$ with the shear viscosity for $\gammap<70$~s$^{-1}$.  We stress again the fact that the flow behavior along the decreasing branch of the flow curve is very complex since widely different local shear rates are evidenced in the viscous zone for the same value of the shear stress.

Finally note that the exact location of the fractures or of the shear bands may depend upon the interactions of the gel with the walls. Experiments dealing with another organogel have shown that the fracture may locate at the wall \cite{Grondin:preprint}. In this case, the gel was not cooled directly in the Couette cell but before being loaded into the cell which clearly limits the links between the gel and the cell.

\section{Summary, discussion and conclusion}

In this work, we have studied the nonlinear rheological properties of an organogel. 
Under small applied shear rates (regime 1), a flowing region is nucleated in the middle of the gap of a Couette cell. Increasing the applied shear rate increases the size of the flowing region. Moreover backscattered ultrasonic signals clearly evidence three-dimensional displacements and fracture-like events in this ``viscous zone.'' 
Time-averaged velocity measurements show that the local flow behavior in the viscous zone is characterized by a decreasing local flow curve and a very large range of accessible shear rates for a given shear stress. At higher applied shear rates (regime 2), an increasing flow curve is recovered but a solid-like region persists close to the stator. When the mechanical stress is released, the fractures heal rapidly and the sample recovers its initial elastic properties. 

We believe that the present work sheds new light on the jamming concept and wish to further discuss our results in this context. Indeed the jamming phase diagram suggested by Liu and Nagel \cite{Liu:98} provides a unified picture of the ergodic-to-nonergodic transition in systems as diverse as 
 clay suspensions \cite{Knaebel:00}, emulsions \cite{Cipelliti:03}, or microgel solutions \cite{Cloitre:00}. This diagram considers the possibility 
 of unjamming the system by increasing its temperature, releasing its density, or increasing the amplitude of some applied shear stress. 
Although the jamming transition presents some universal features (universal slow dynamics  and ageing behavior have been observed in a wide variety of jammed systems such as colloidal suspensions, concentrated emulsions, and lamellar gels), we argue that
the unjamming mechanisms are not universal. In a previous work, we have pointed out that the flow of concentrated emulsions in the vicinity of the yield stress may be homogeneous or heteregeneous depending on the interactions between the droplets: shear banded flow occurs only for adhesive emulsions whereas the flow of nonadhesive emulsions remains homogeneous through the yielding transition \cite{Becu:06}.

With the present work, we bring new elements that corroborate this point of view. 
First we have found that the viscous (or ``unjammed'') zone in regime 1 is nucleated in the middle of the gap of a Couette cell. Such a location for the flowing material had not been reported before: previous works have shown that, in the vicinity of the yield stress, flow is usually initiated in the region of higher shear stress {\it i.e.}, close to the inner cylinder of a Couette cell \cite{Coussot:02}. Moreover for shear rates in regime 1, the flow in the viscous zone is quite unusual with three-dimensional features and fracture-like events. Rotational motions of the aggregates and fractures induce a loss of energy. These events decrease in number and in amplitude as the applied shear rate is increased. Thus a lower shear stress is required to maintain the flow at higher shear rates. This scenario could provide an explanation for the decreasing branch observed in regime 1 on the local flow curve. We also point out that the jammed phase and the unjammed one may coexist over a very large range of shear stresses. In this case, defining and measuring a ``yield stress'' becomes very difficult. For shear rates in regime 2, fractures and three-dimensional motions disappear. Although the global flow curve points to a low viscosity liquid behavior, local measurements show that the system remains jammed close to the stator over a significant thickness and for the highest investigated shear rates. In the framework of the jamming phase diagram, this last feature remains rather puzzling.

This study clearly shows that local measurements are required to understand the flow of complex fluids such as organogels. It also demonstrates that complex three-dimensional flows may occur in viscous systems, even at very low shear rates. In particular, although restricted to a specific organogel with large-scale heterogeneities, the present work reveals the existence of both ``fast'' and ``slow'' spatially localized events involved in the unjamming transition. Instead of claiming for any generality, we rather wish to emphasize that yielding mechanisms may well be widely different from one material to another. In our opinion, investigating more deeply the nature of the interplay between the complex local flow behavior observed along the decreasing branch of the flow curve and the three-dimensional motions and fracture-like events revealed by the ultrasonic BS signals constitutes the most promising research direction on such thixotropic systems driven far from equilibrium.

\begin{acknowledgments}
The authors thank L. Bocquet, R. Cl\'erac, O. Roubeau, and J.-B. Salmon for fruitful discussions.
\end{acknowledgments}

\end{document}